\documentclass[12pt]{article}
\usepackage[dvips]{graphicx}
\usepackage{epsfig,cite}
\usepackage {amsmath}
\usepackage{amssymb}
\include{epsf}
\newcount\timecount
\newcount\hours \newcount\minutes  \newcount\temp \newcount\pmhours
\hours = \time
\divide\hours by 60
\temp = \hours
\multiply\temp by 60
\minutes = \time
\advance\minutes by -\temp
\def\hour{\the\hours}
\def\minute{\ifnum\minutes<10 0\the\minutes
            \else\the\minutes\fi}
\def\clock{
\ifnum\hours=0 12:\minute\ AM
\else\ifnum\hours<12 \hour:\minute\ AM
      \else\ifnum\hours=12 12:\minute\ PM
            \else\ifnum\hours>12
                 \pmhours=\hours
                 \advance\pmhours by -12
                 \the\pmhours:\minute\ PM
                 \fi
            \fi
      \fi
\fi
}

\def\monthname{\relax\ifcase\month 0/\or January\or February\or
   March\or April\or May\or June\or July\or August\or September\or
   October\or November\or December\else\number\month/\fi}

\def\bold#1{\setbox0=\hbox{$#1$}%
     \kern-.025em\copy0\kern-\wd0
     \kern.05em\copy0\kern-\wd0
     \kern-.025em\raise.0433em\box0 }

\def\beq{\begin{equation}}
\def\eeq{\end{equation}}

\def\ga{\mathrel{\raise.3ex\hbox{$>$\kern-.75em\lower1ex\hbox{$\sim$}}}}
\def\la{\mathrel{\raise.3ex\hbox{$<$\kern-.75em\lower1ex\hbox{$\sim$}}}}
\def\gev{{\rm \, Ge\kern-0.125em V}}
\def\tev{{\rm \, Te\kern-0.125em V}}
\def\gyr{{\rm \, G\kern-0.125em yr}}

\def\gappeq{\mathrel{\rlap {\raise.5ex\hbox{$>$}}
{\lower.5ex\hbox{$\sim$}}}}
\def\lappeq{\mathrel{\rlap{\raise.5ex\hbox{$<$}}
{\lower.5ex\hbox{$\sim$}}}}
\def\Toprel#1\over#2{\mathrel{\mathop{#2}\limits^{#1}}}

\def\stau{\widetilde \tau}

\def\mchi{m_{\tilde \chi}}

\def\m12{m_{1\!/2}}

\newcommand\iso[2]{\mbox{${}^{#2}${\rm #1}}}
\def\he#1{\iso{He}{#1}}
\def\be#1{\iso{Be}{#1}}
\def\li#1{\iso{Li}{#1}}
\def\b1#1{\iso{B}{1#1}}

\def\stau{\tilde{\tau}}

\def\mgrav{m_{3/2}}
\def\grav{\widetilde{G}}
\def\qbar{\overline{q}}

\def\mchi{m_{\chi}}

\def\bea{\begin{eqnarray}}
\def\eea{\end{eqnarray}}

\def\nrg{\epsilon}
\def\pd{\partial}

\def\pref#1{(\ref{#1})}
\def\beqar{\begin{eqnarray}}
\def\eeqar{\end{eqnarray}}

\begin{document}

\begin{titlepage}
\pagestyle{empty}
\baselineskip=21pt
\rightline{\tt astro-ph/0608562}
\rightline{CERN--PH--TH/2006-168, UMN--TH--2516/06, FTPI--MINN--06/29}
\vskip 2mm
\begin{center}
{\large {\bf   Bound-State Effects on Light-Element Abundances in Gravitino Dark Matter Scenarios}}

\end{center}
\begin{center}
\vskip 2mm
{\bf Richard~H.~Cyburt}$^1$, {\bf John~Ellis}$^2$, 
{\bf Brian~D.~Fields}$^{3}$, \\
{\bf Keith~A.~Olive}$^{4}$, and {\bf Vassilis~C.~Spanos}$^{4}$
\vskip 0.1in
{\it
$^1${TRIUMF, Vancouver, BC V6T 2A3 Canada}\\
$^2${TH Division, PH Department, CERN, 1211 Geneva 23, Switzerland}\\
$^3${Departments of Astronomy and of Physics, University of Illinois, Urbana, IL 61801, USA}\\
$^4${William I. Fine Theoretical Physics Institute, \\
University of Minnesota, Minneapolis, MN 55455, USA}}\\
\vskip 3mm
{\bf Abstract}
\end{center}
\baselineskip=18pt \noindent

If the gravitino is the lightest supersymmetric particle and the
long-lived next-to-lightest sparticle (NSP) is the stau, the charged
partner of the tau lepton, it may be metastable and form bound states with
several nuclei. These bound states may affect the cosmological abundances of \li6 and
\li7 by enhancing nuclear rates that would otherwise be strongly
suppressed. We consider the effects of these enhanced rates on the final
abundances produced in Big-Bang nucleosynthesis (BBN), including
injections of both electromagnetic and hadronic energy during and after
BBN. We calculate the dominant two- and three-body decays of both
neutralino and stau NSPs, and model the electromagnetic and hadronic decay
products using the {\small PYTHIA} event generator and a cascade equation.  
Generically, the introduction of bound states drives
light element abundances further from their observed values;
however, for small regions of parameter space bound state effects
can bring lithium abundances in particular in better accord
with observations.
We show that in regions where the stau is the NSP with a lifetime longer
than $10^3-10^4$~s, the abundances of \li6 and \li7 are far in excess of
those allowed by observations. For shorter lifetimes of order $1000$~s, we
comment on the possibility in minimal supersymmetric and supergravity
models that stau decays could reduce the \li7 abundance from standard BBN
values while at the same time enhancing the \li6 abundance.

\vspace*{5mm}
\vfill
\leftline{CERN--PH--TH/2006-168}
\leftline{August 2006}
\end{titlepage}
\baselineskip=18pt

\section{Introduction}

The abundances of the light nuclei produced by primordial Big-Bang
nucleosynthesis (BBN) provide some of the most stringent constraints on
the decays of unstable massive particles during the early
Universe~\cite{holtmann, kkm, kohri, cefo, grant, kkm2, kmy, Jedamzik:2006xz,cefos2}. 
This is because the
astrophysical determinations of the abundances of deuterium (D) and \he4  agree
well with those predicted by homogeneous BBN calculations, and also the
baryon-to-photon ratio $\eta \equiv n_b/n_\gamma \propto \Omega_b h^2$ 
needed for the success of these
calculations~\cite{cfo1,bbn2} agrees very well with that
inferred~\cite{cfo2} from observations of the power spectrum of
fluctuations in the cosmic microwave background (CMB). The value of $\eta
= (6.11 \pm 0.25) \times 10^{-10}$
that they indicate~\cite{wmap} is now quite precise, reducing
one of the principal uncertainties in the previous BBN calculations.

However, it is still difficult to reconcile the BBN predictions for the
lithium isotope abundances with observational indications on the
primordial abundances.  The discovery of the ``Spite''
plateau~\cite{spite}, which demonstrates a near-independence of the \li7
abundance from the metallicity in Population-II stars, suggests a
primordial abundance in the range ${\rm \li7/H} \sim (1 - 2) \times
10^{-10} $ ~\cite{rbofn}, whereas standard BBN with the CMB value of
$\eta$ would predict ${\rm \li7/H} \sim 4 \times 10^{-10}
$~\cite{cfo1,bbn2}. In the case of $\li6$, the data \cite{li6obs} lie a
factor $\sim 1000$ above the BBN predictions~\cite{bbnli6}, and fail to
exhibit the dependence on metallicity expected in models based on
nucleosynthesis by Galactic cosmic rays~\cite{li6cr}. On the other hand,
the $\li6$ abundance may be explained by pre-Galactic Population-III
stars, without additional over-production of $\li7$ \cite{rvo}.

The concordance between BBN predictions and the observed abundances of D
and \he4 is relatively fragile and could have been upset by decays of
massive unstable particles.  Electromagnetic and hadronic showers produced
in decays occurring during or shortly after BBN induce new reactions which
may either create or destroy light nuclides. Concrete examples of unstable
but long-lived particles are found in supersymmetric theories. In
particular, models based on the constrained version of the minimal
$R$-parity conserving supersymmetric standard model (CMSSM) with a
gravitino as the lightest supersymmetric particle (LSP) have been
considered in this context~\cite{SFT,eoss5,vcmssm,EOV,Jed1,Jed2}. In these
models with a gravitino LSP, the next-to-lightest supersymmetric particle
(NSP) may be either the lightest neutralino $\chi$ or the lighter stau
${\tilde \tau_1}$, and the decay lifetime of the NSP can range from
seconds to years depending on the specific model and gravitino mass.  Our
previous results concentrated on relatively long lifetimes ($\tau >
10^4$~s), and the effects of the electromagnetic showers on the
light-element abundances~\cite{cefo,eoss5,vcmssm,EOV}.

The effects of hadronic injections due to late decays of the NSP during
BBN have also been studied extensively by other
authors~\cite{kohri,kkm2,kmy,karsten,jed,stef}. In particular, it has been
shown for relatively short lifetimes of order $10^3$~s that decays may
simultaneously increase the \li6 abundance and decrease the \li7
abundance~\cite{karsten,jed}.  On the other hand, it has been
shown~\cite{EOV} that purely electromagnetic showers cannot reduce the
\li7 abundance sufficiently without also overproducing D relative to
\he3~\cite{sigl}. In~\cite{SFT} the authors have calculated the principal
three-body decays of a stau NSP and, using the results of the
analysis~\cite{kkm2} have explored regions where the \li7 puzzle can be
solved in an unconstrained supersymmetric model.  The authors
of~\cite{kkm2} used the {\small PYTHIA}~\cite{pythia} model for $e^+e^-$
annihilation to hadrons in order to simulate the hadronic decays of the
unstable particle. An analogous simulation was used in~\cite{Jed1}, both
to locate regions in the parameter space of the CMSSM which are compatible
with the BBN constraints, and to solve the lithium problem~\cite{Jed2}.

It has recently been pointed out that, if it has electric charge, the NSP
forms bound states with several nuclei \cite{maxim}.
Due to the large NSP mass ($m_{\rm nsp} \gg m_{\rm nucleon}$),
the Bohr radii of these bound states 
$\sim \alpha^{-1} m_{\rm nucleon}^{-1} \sim 1 \ {\rm fm}$
are of order the nuclear size. Consequently, 
nuclear reactions with nuclei in bound states are catalyzed,
due to partial screening of the Coulomb barrier, and
due to the opening of virtual photon channels in radiative capture
reactions.  Ref.~\cite{maxim} used analytic approximations to
argue that the $d(\alpha,\gamma)\li6$ reaction is
enhanced by an enormous factor $\sim 10^6$, leading
to \li6 production far beyond tolerable levels
over large regions of parameter space.
Other effects of bound states were considered in~\cite{kota,kapling}.

Here, we present results
from a new BBN code that includes the nuclear reactions induced by
hadronic and electromagnetic showers generated by late gravitational decays of the NSP,
together with the familiar network of
nuclear reactions used to calculate the primordial abundances of the light
elements Deuterium (D), \he3, \he4 and \li7.
In addition, we include the effects of the bound states
when the decaying particle is charged.  As claimed in~\cite{maxim},
these bound states lead to large enhancements in otherwise heavily suppressed rates.  
We calculate the abundances of the various bound states and 
include both the Coulomb enhancement as well as virtual photon effects
in radiative capture reactions.

We use as frameworks for this study both the CMSSM and mSUGRA
models~\cite{vcmssm}, where the NSP could be either the lighter stau or
the lightest neutralino.  We have calculated the dominant two- and
three-body gravitational decays of these sparticles and use the {\small
PYTHIA} Monte Carlo event generator and a cascade equation to model the
resulting electromagnetic (EM) and hadronic (HD)  spectra for each point
of the parameter space of the supersymmetric model.  The resulting
accurate determinations of the abundances of the light elements, as
altered by these late injections and the bound-state effects, enable us to
delineate regions of the parameter space of the supersymmetric models
which are compatible with the BBN constraints. In addition, we look for
regions where the \li6 and \li7 puzzles can be solved in the context of
these supersymmetric models. We find that for lifetimes $\tau < 10^3 - 10^4$~s, the
enhanced rates of \li6 and \li7 production, exclude gravitino dark matter
(GDM) with a stau NSP.  At smaller lifetimes, we see that it is the \li7
destruction rates which are enhanced, facilitating a solution to the Li
problems.

\section{Electromagnetic and Hadronic NSP Decays}
\label{sec:decays}

The CMSSM is determined by four real parameters, namely
the soft supersymmetry-breaking scalar mass $m_0$,  the gaugino mass 
$m_{1/2}$, the trilinear coupling  $A_0$ (each taken to be universal at the Grand Unification scale),
the ratio of Higgs vevs $\tan \beta$, and the 
sign of the $\mu$ parameter.  Here, for simplicity, we restrict our attention to $A_0 = 0$ and $\mu > 0$.
The mass of the gravitino is also a free parameter in the CMSSM, and if is chosen
to be less than $\min(m_\chi,m_{\tilde \tau})$ the resultant model has GDM. 
In mSUGRA models~\cite{vcmssm}, $\tan \beta$ is no longer a free parameter, nor
is the gravitino mass which is now equal to $m_0$ at the Grand Unification scale.
GDM is a inevitable consequence in mSUGRA models when $m_0$ is relatively small.  
In this case, the NSP is typically the stau, but it may also be the neutralino if $A_0/m_0 \la 1.7$.
The abundances of light elements provide some of the most important
constraints on this gravitino LSP scenario~\cite{SFT,eoss5,vcmssm,EOV,Jed1,Jed2}.  
They also impose important constraints on the neutralino LSP scenario,
since a gravitino NSP would also decay slowly.  Here, however, 
we restrict our attention to GDM scenarios with either a stau or neutralino NSP.

In order to estimate the lifetime of the NSP, as well as the various branching ratios
and the resulting EM and HD spectra, one must calculate the partial widths 
of the dominant relevant decay channels of the NSP~\footnote{Analytical 
results for all of the relevant partial widths will be presented elsewhere~\cite{cefos2}.}.
The decay products that yield EM energy obviously include directly-produced photons,
and also indirectly-produced photons, charged leptons (electrons and muons) which are produced via the secondary decays of gauge and Higgs bosons, as well as neutral pions ($\pi^0$).
Hadrons (nucleons and  mesons such as the $K_L^0$, $K^\pm$ and $\pi^\pm $) 
are usually produced through the secondary decays of gauge and Higgs bosons,
as well (for the mesons) as via the decays of the heavy $\tau$ lepton.
It is important to note that mesons decay before interacting with the 
hadronic background~\cite{kohri,SFT}. Hence they are irrelevant 
to the BBN processes and to our analysis, except via their decays into photons
and charged leptons. Therefore, the HD injections on which we focus our attention are those that 
produce nucleons, namely the decays via gauge and Higgs bosons and quark-antiquark pairs.

For the neutralino NSP $\chi$, we include the two-body decay
channels $\chi \to \grav \, H_i$ and $\chi \to \grav \, V$, where $H_i=h,H,A$ and $V=\gamma,Z$. 
These are the dominant gravitational  decays of $\chi$, whose analytical expressions
have been presented in~\cite{eoss5}. In addition, we include here the dominant 
three-body decays  $\chi \to \grav \, \gamma^*  \to \grav \, q  \qbar$,  
 $\chi \to \grav \, \gamma^* \to \grav \, W^+ W^-$, 
$\chi \to \grav \, W^+ W^-$ and the corresponding interference terms.
In general, the two-body channel $\chi \to \grav \, \gamma$ dominates the 
$\chi$ NSP decays and yields the bulk of the injected EM energy.
When the $\chi$ is heavy enough to produce a real $Z$ boson, the next most important 
channel is $\chi \to \grav \, Z$, which is also the dominant channel for producing HD 
injections in this case. The Higgs boson channels are smaller by a few orders
of magnitude, and those to heavy Higgs bosons ($H,A$) in particular
become kinematically accessible only for heavy $\chi$ in the large-$m_{1/2}$ region.  
Turning to the three-body channels, the decay through the virtual photon to a $q \qbar$ pair 
can become comparable to the subdominant channel $\chi \to \grav \, Z$, 
injecting nucleons even in the kinematical region $\mchi < \mgrav + M_Z$,
where direct on-shell $Z$-boson production is not possible~\footnote{In principle, one 
should also include $q \qbar$ pair production through the virtual $Z$-boson
channel $\chi \to \grav \, Z^*  \to \grav \, q  \qbar$~\cite{kmy} 
and the corresponding interference term. However,
this process is suppressed by a factor of $M_Z^4$ with respect to
$\chi \to \grav \, \gamma^*  \to \grav \, q  \qbar$, and the interference term is also
suppressed by $M_Z^2$. Numerically, these contributions are
unimportant, and therefore we drop these amplitudes in our calculation.}. 
Finally, we note that the three-body decays to $W^+W^-$ pairs and a gravitino are usually 
at least five  orders of magnitude smaller. 

\begin{figure}
\vskip 0.5in
\vspace*{-0.75in}
\begin{minipage}{8in}
\epsfig{file=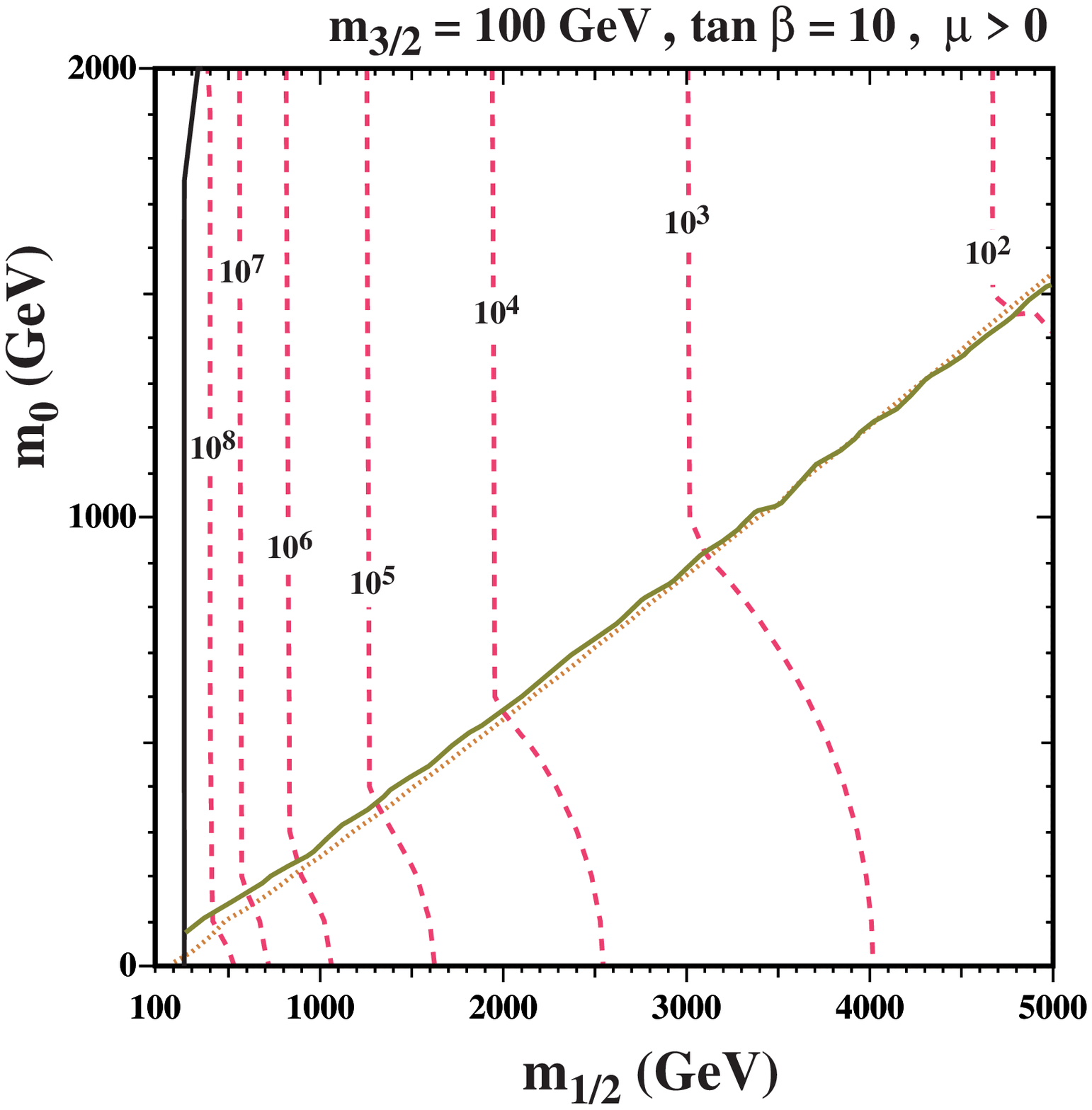,height=3.3in}
\hspace*{-0.17in}
\epsfig{file=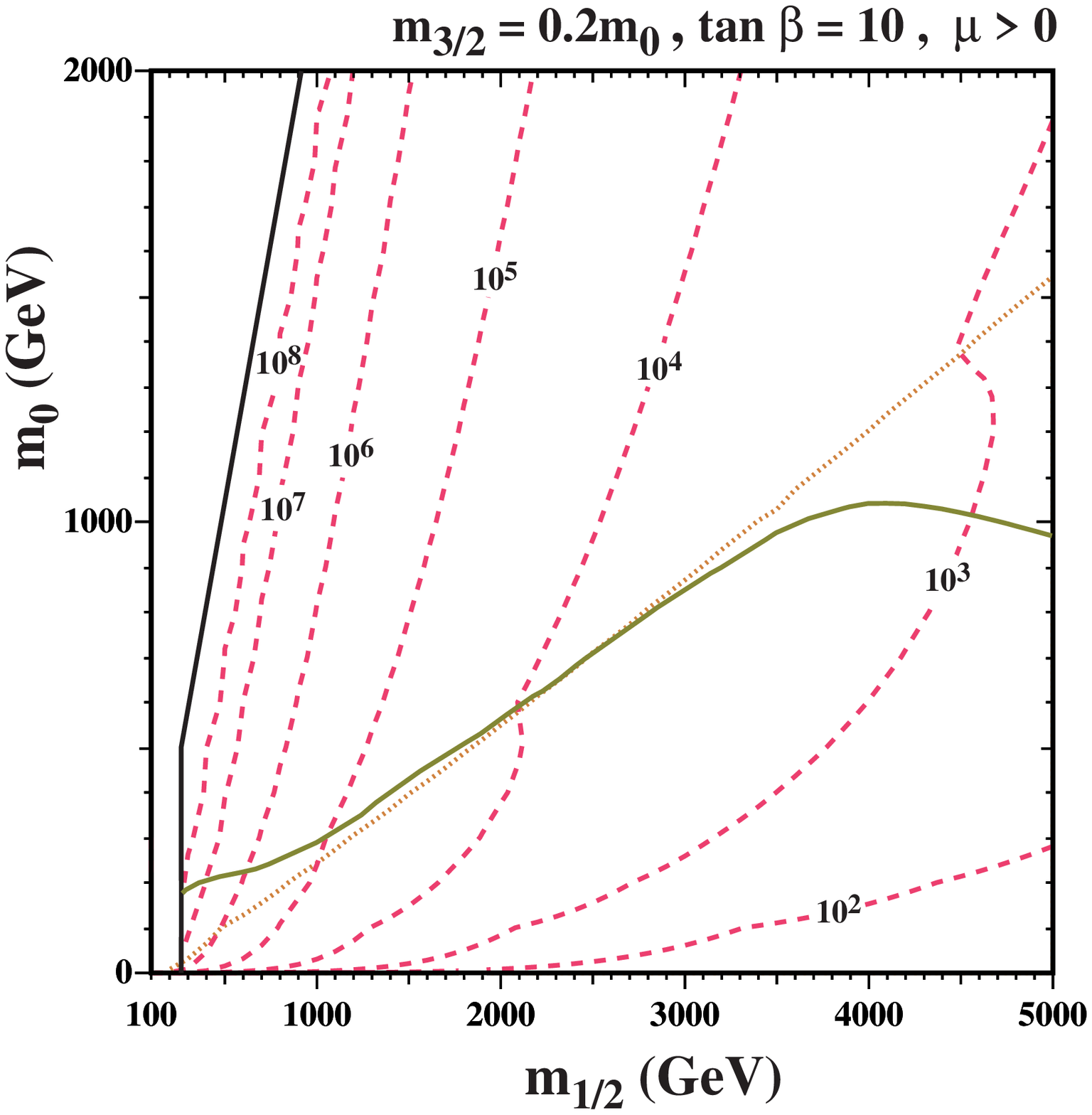,height=3.3in}
\hfill
\end{minipage}
\begin{minipage}{8in}
\epsfig{file=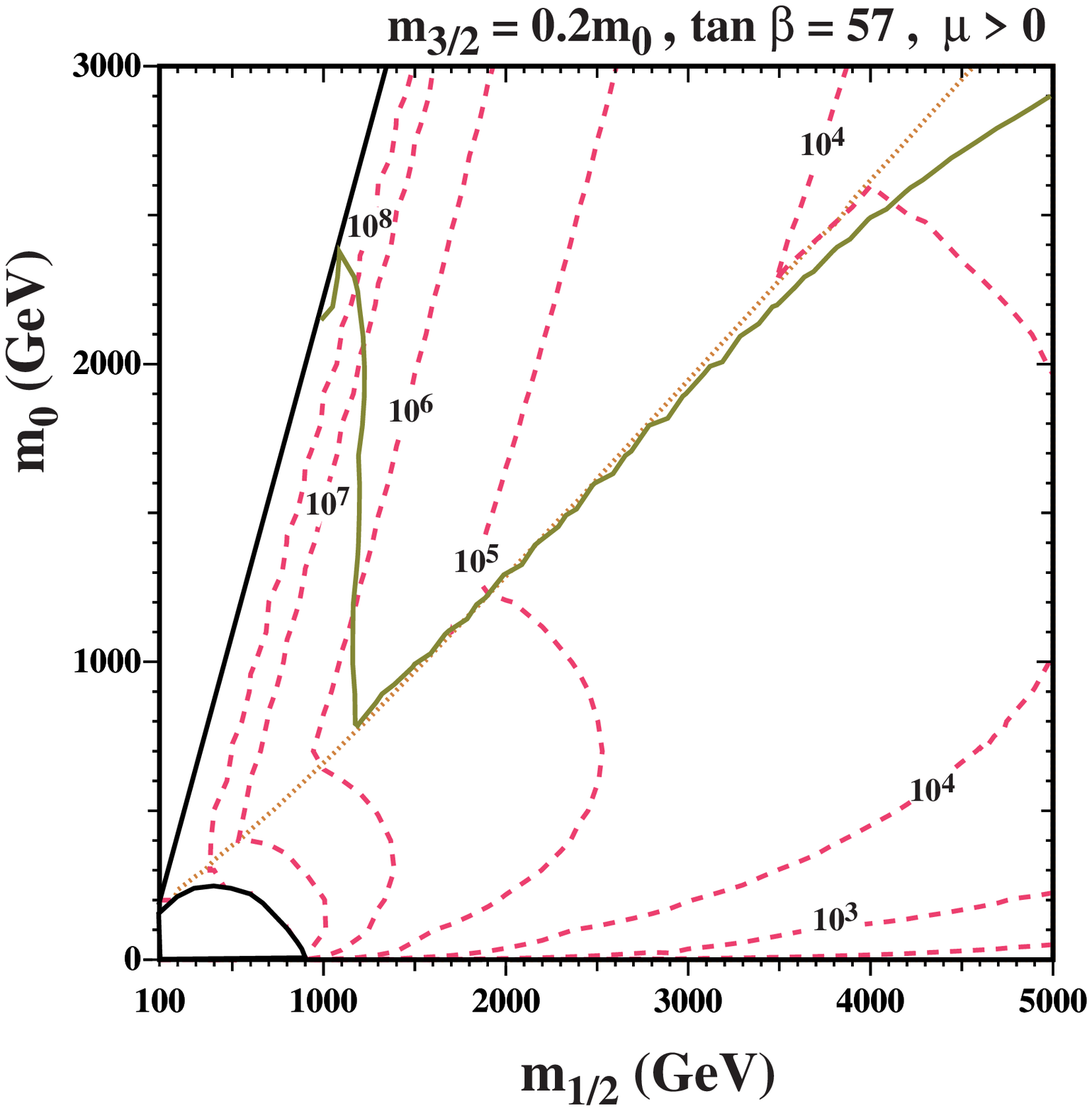,height=3.3in}
\hspace*{-0.2in}
\epsfig{file=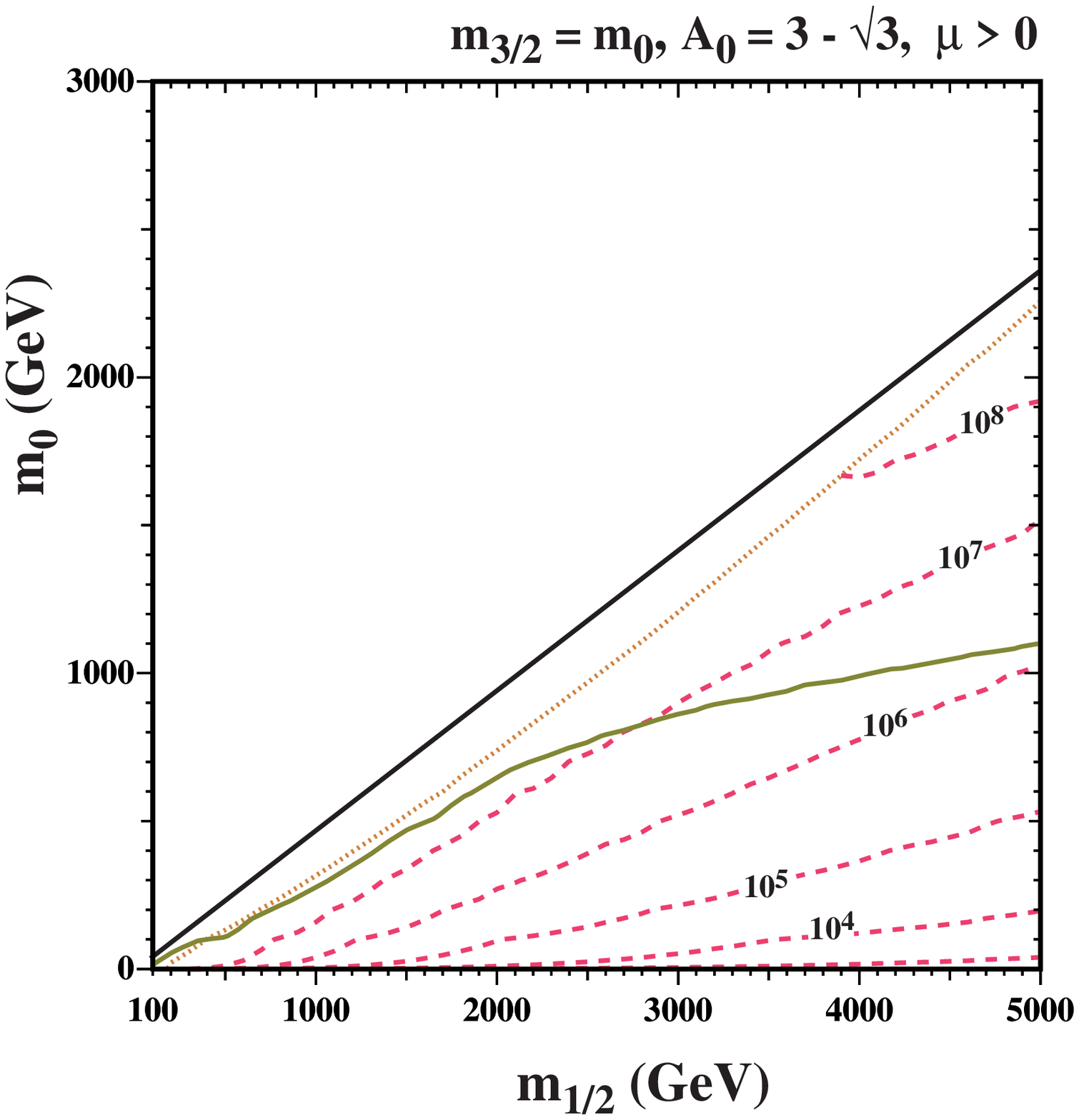,height=3.3in} \hfill
\end{minipage}
\caption{
{\it 
The NSP lifetime contours (in seconds) for the supersymmetric models 
discussed in Figs.~\protect\ref{fig:CMSSM10} and 
\protect\ref{fig:CMSSM57}.}}
\label{fig:lifetimes} 
\end{figure}

For general orientation, we present in Fig.~\ref{fig:lifetimes} contours
(in seconds) for the supersymmetric models discussed later and presented
in Figs.~\ref{fig:CMSSM10} and \ref{fig:CMSSM57}. We display the
$m_{1/2},m_0$ planes for variants of the CMSSM with different gravitino
masses for $\tan \beta = 10, A_0 = 0$ (panels a and b), and $\tan \beta =
57, A_0 = 0$ (panel c). Panel d displays an analogous plane for a mSUGRA
model with $A_0/m_0 = 3 - \sqrt{3}$ at the GUT scale, in which $\tan
\beta$ is determined at each point by the electroweak vacuum conditions.
In addition to the NSP lifetime contours (labelled by the lifetime in
seconds), we show the boundaries between regions with neutralino and stau
NSPs (dotted red lines) and the upper limit on the gravitino mass density
(solid brown lines). These curves are also found in
Figs.~\ref{fig:CMSSM10} and \ref{fig:CMSSM57}.

Having calculated the partial decay widths and branching ratios, we employ
the {\small PYTHIA} event generator~\cite{pythia} to model both the EM and
the HD decays of the direct products of the $\chi$ decays. We first
generate a sufficient number of spectra for the secondary decays of the
gauge and Higgs bosons and the quark pairs.  Then, we perform fits to
obtain the relation between the energy of the decaying particle and the
quantity that characterizes the hadronic spectrum, namely $dN_h/dE_h$, the
number of produced nucleons as a function of the nucleon energy.  These
spectra and the fraction of the energy of the decaying particle that is
injected as EM energy are then used to calculate the light-element
abundances.

An analogous procedure is followed for the $\stau$ NSP case. As the
lighter stau is predominantly right-handed, its interactions with $W$
bosons are very weak (suppressed by powers of $m_\tau$) and can be
ignored. The decay rate for the dominant two-body decay channel, namely
$\stau \to \grav \, \tau$, has been given in~\cite{eoss5}. However, this
decay channel {\em does not yield any nucleons}.  Therefore, one must
calculate some three-body decays of the $\stau$ to obtain any protons or
neutrons. The most relevant channels are $\stau \to \grav \, \tau^* \to
\grav \, Z \, \tau $, $\stau \to Z \, \stau^* \to \grav \, Z \, \tau $,
$\stau \to \tau \chi^* \to \grav \, Z \, \tau $ and $\stau \to \grav \, Z
\, \tau $~\cite{SFT}.  We calculate these partial widths, 
and then use {\small PYTHIA} to obtain the
hadronic spectra and the EM energy injected by the secondary $Z$-boson and
$\tau$-lepton decays. As in the case of the $\chi$ NSP, this information
is then used for the BBN calculation.

We stress that this procedure is repeated separately for each point in the
supersymmetric parameter space sampled. That is, given a set of parameters
$m_0, m_{1/2}, A_0, \tan \beta$, sgn$(\mu)$, and $m_{3/2}$, once the
sparticle spectrum is determined, all of the relevant branching fractions
are computed, and the hadronic spectra and the injected EM energy
determined case by case. For this reason, we do not use a global parameter
such as the hadronic branching fraction, $B_h$, often used in the
literature. In our analysis, $B_h$ is computed and differs at each point
in the parameter space.

\section{Electromagnetic and Hadronic Showers 
During Primordial Nucleosynthesis}

The dominant effect of hadronic decays of the NSP during  BBN
is the addition of new interactions between
hadronic shower particles and background nuclides~\footnote{The 
decays of the NSP affect, in principle,
the expansion rate of the Universe. However, this effect is
negligible for NSP abundances low enough to respect the
other constraints discussed below.}.
These alter the evolution and final values of
the light-element abundances, as follows.
For each background species $i$, 
let the rate of interactions of decay hadrons
per background particle be $\Gamma_i$.
Then the  $i$ abundance per background baryon, $Y_i = n_i/n_B$,
changes according to
\beq
\label{eq:bbnrate}
\pd_t Y_i =  \left( \pd_t Y_i \right)_{\rm SBBN}
 + \left( \pd_t Y_i \right)_{\rm HAD} 
+ \left( \pd_t Y_i \right)_{\rm EM},
\eeq
where $\dot{Y}_{\rm SBBN}$ gives the rate of change of the $i$
abundance in standard
BBN.  We have also included in \pref{eq:bbnrate} the effects of
electromagnetic interactions due to NSP decays, either from the decays directly
to photons or leptons, or through electromagnetic
secondaries in the hadronic showers.  These are treated as in~\cite{cefo},
but are not dominant when hadronic branchings are significant.  
All we need to know is the total EM energy released per
decay in any given channel, which may become more complicated in the
three-body case, but can easily be calculated.

Including hadronic decays in BBN thus
amounts to an evaluation of
the interaction rates
\beq
\label{eq:dYdthad}
\left( \pd_t Y_i \right)_{\rm HAD} 
= - \Gamma_{\rm i \rightarrow {\rm inel}} Y_i 
  + \sum_{hb} \Gamma_{hb \rightarrow i} Y_b.
\eeq
The first term
accounts for $i$ destruction by hadro-dissociation,
where $\Gamma_{\rm i \rightarrow {\rm inel}}$ is the
total rate (for a given species $i$) of all inelastic interactions of shower
particles with $i$.
The second term accounts for
production via the hadro-dissociation of
heavier background species, e.g.,
$p_{\rm shower} \alpha_{\rm bg} \rightarrow d$.
The sum runs over shower species $h$
and background targets $b$. 
In the case of the lithium isotopes, production also
occurs via the interactions of energetic (i.e., nonthermal) mass-3 dissociation
products with background \he4,
e.g., $\he3 \alpha \rightarrow \li{6,7} + \cdots$. 

Consider a nonthermal  hadronic projectile species $h$,
with energy spectrum $N_{h}(\nrg,t) = dN_h/dVd\nrg$
and total number density $n_h = \int N_{h} d\nrg$.
The rate for $i$ production due to $hb \rightarrow i$
is
\beq
\Gamma_{hb \rightarrow i}
   =  \int N_{h}(\nrg,t) \sigma_{hb \rightarrow i}(\nrg)  d\nrg .
\eeq
The rates $\Gamma_i$ thus depend on 
the decay particle, on
the background abundances,
and most importantly from the point of view of
implementation, on the shower development
and evolution of $N_{h,\nrg}$
in the background environment.


We wish to follow the evolution of $N_{h}$
over the multiple shower generations produced by the 
initial hadronic NSP decay products.
In the context of BBN,
this problem of shower development has
been approached via direct computation of
the multiple generations of shower particles~\cite{kohri,kkm,karsten}.
In this approach, the final particle spectrum is obtained
by iterating an initial decay spectrum,
accounting for both the energy losses and
the energy distributions of collision products.

We introduce here an equivalent alternative
approach, based on a cascade equation, emulating
the well-studied
treatment of hadronic 
shower development due to cosmic-ray interactions
in the atmosphere.
The spectrum of $h$ evolves according to
\beq
\label{eq:cascade}
\pd_t N_{h}(\epsilon)
  =  J_{h}(\epsilon) - \Lambda_{h}(\epsilon)N_{h}(\epsilon)-\pd_\nrg\left[b_{h}(\epsilon) N_{h}(\epsilon)\right], 
\eeq
where $J_{h}$ is the sum of all source terms, $\Lambda_{h}$ is the sum
of all sink terms, and $b_h$ is the energy-loss rate of particle-conserving processes.  
The energy-loss term is assumed to be the dominant process of energy
transfer, in which case tertiary processes are limited to down-scattering.
The sink term has two contributions, due to elastic and inelastic scattering;
the source term has three contributions, due to direct injection, elastic
down-scattering and inelastic down-scattering.

Each nonthermal species
$i$ evolves according to a cascade equation
of the form \pref{eq:cascade}, and together these constitute a coupled set of
equations.
Because the source term includes the elastic term with
$N_{i}$ inside the integral, these equations
are of integro-differential form.  The
integration is therefore not immediate.
Previous work on hadronic decays has adopted
a Monte Carlo approach to the solution;
our method is to solve the differential equation.

We note that because this is an integral equation, 
we can adopt an iterative approach to the solution.
To make our initial guess, we ignore the downscattering
and solve for $N_h$ with decays being the only
source.  The solution can then 
be written in terms of the following quadrature
\beq
N_h^{(i)}(\nrg,t) = \frac{1}{b(\nrg)} \int_\nrg^\infty 
   d\nrg^\prime \ J_h^{(i)}(\nrg^\prime,t) \ e^{-R(\nrg^\prime,t)} ,
\eeq
where our initial guess takes $J_h^{(0)} = q_X(\nrg)$ only.
Here the exponential ``optical depth'' factor
\beq
R(\nrg^\prime,t) 
 = \int_\nrg^{\nrg^\prime} d \nrg^{\prime \prime} \  
    \frac{\Gamma(\nrg^{\prime \prime},t)}{b(\nrg^{\prime\prime})}
\eeq
is a measure of the average number of 
inelastic interactions over the time taken to
lose energy electromagnetically from $\nrg^\prime$ to $\nrg$.

The full cascades can then be treated iteratively,
correcting the approximation to include the 
redistribution and production of nucleons in
scattering events.
We do this
by using the previous solution $N_h^{(i)}$ to update the 
source term  $J_h^{(i+1)}$ by including
the downscattering terms.
These distributions converge after a few iterations.
We then insert them into (\ref{eq:dYdthad}) and solve for the
hadro-dissociation rate.  This iterative procedure is similar to what
previous studies have done.  However, rather than including the exponent in the
integral, they treat the exponential $R$ term as a delta function,
evaluated at $\nrg_*$ ($R(\nrg_*,\nrg,t)=1$).  

Full details of our method will be given in \cite{cefos2}.

\section{Bound-State Effects}
\label{sec:bs}

It has recently been pointed out~\cite{maxim} that the presence of a charged
particle, such as the stau, during BBN can alter the light-element abundances in a significant way
due to the formation negatively-charged staus of  bound states (BS)  with charged nuclei.
The binding energies of these states are 
$\alpha^2 Z_i^2 m_i/2 \approx 30 Z_i^2 A_i \ {\rm keV}$,
and the Bohr radii $\sim (\alpha Z_i m_i) \sim 1 \ Z_i^{-1} A_i^{-1} \ {\rm fm}$.
For species such as \he4, \li7 and \be7,
these energy and length scales are close to those of nuclear
interactions, and it thus turns out that bound state formation
results in catalysis of nuclear rates via two mechanisms.

One immediate
consequence of the bound states is a reduction of
the Coulomb barrier for nuclear reactions,
due to partial screening by the stau.
Since Coulomb repulsion dominates the charged-particle rates,
all such rates are enhanced.  Specifically, for the case of 
an initial state $A_1 + A_2$,
Coulomb effects lead to a exponential suppression via a
penetration factor which scales as $Z_1^{2/3} Z_2^{2/3} A^{1/3}$,
with $A = A_1 A_2/(A_1+A_2)$.
Introduction of a bound state $(\stau,A_2)$ decreases the
target charge to $Z_2-1$ and the system's reduced mass 
number to $A = A_1$; both effects lower the Coulomb suppression.
We include these effects for all reactions with bound states.

\begin{table}[h]
\caption{\it
Bound-state virtual photon enhancements to radiative capture cross sections for various reactions.
The third (fourth) column is the threshold energy for the standard (catalyzed)  BBN.
The last column is the virtual photon enhancement factor of the catalyzed cross section relative to standard BBN.
 }
 
\label{tab:enhance}

\begin{center}
\begin{tabular}{ccccc} 
\hline\hline
 & EM & $A(B,\gamma)$ & $[X^- A](B,C)X^-$ & Enhancement \\
Reaction  &  Transition  & $Q_{\rm SBBN}$ (MeV) &  $Q_{\rm CBBN}$ (MeV) & $\sigma_{\rm CBBN}/\sigma_{\rm SBBN}$  \\ \hline\hline
$d(\alpha,\gamma$)\li6     &  E2 & 1.474   & 1.124  & 7.0$\times10^7$ \\ 
\iso{H}{3}($\alpha,\gamma$)\li7     &  E1 & 2.467   & 2.117  & 1.0$\times10^5$ \\ 
\he3($\alpha,\gamma$)\be7  &  E1 & 1.587   & 1.237  & 2.9$\times10^5$ \\ 
\li6($p,\gamma$)\be7     &  E1 & 5.606   & 4.716  & 2.9$\times10^4$ \\ 
\li7($p,\gamma$)\be8     &  E1 & 17.255  & 16.325 & 2.6$\times10^3$ \\ 
\be7($p,\gamma$)\iso{B}{8}      &  E1 & 0.137   & -1.323 & N/A \\ \hline
\end{tabular}
\end{center}
\end{table}

An additional effect enhances radiative capture channels $A_2(A_1,\gamma)X$ by
introducing photonless final states in which the stau
carries off the reaction energy transmitted via virtual photon processes.
In particular, the $\he4 (d,\gamma)\li6$ reaction, which is suppressed in 
standard BBN, is enhanced by many orders of magnitude by the presence of the bound states,
as described in~\cite{maxim}. 
Large enhancements of this type affect other 
radiative capture reactions, notably mass-7 production reactions
such as $\he3(\alpha,\gamma)\be7$ and destruction reactions
such as $\li7(\alpha,\gamma)\iso{B}{11}$.
We have included these as well: the corresponding enhancement factors appear in
Table \ref{tab:enhance}.

Bound-state formation and reaction catalysis occurs late in BBN.
The binding energy $E_{\textrm{bin}}$
for the [$\stau,\he4$] bound states is 311 keV,
for [$\stau,\li7$]   952 keV, 
and for the  [$\stau,\be7$]   1490  keV. 
The latter are quite high, of order nuclear binding
energies, and indeed the large \be7 binding 
plays an important role in forbidding \be7 destruction channels
that otherwise would be energetically allowed.
The capture processes
that form these bound states
 typically become effective for temperatures $T_c \approx E_{\textrm{bin}}/30$;
this means that \be7 states form prior to \li7 states, with \he4 states forming last.
At these low temperatures one can ignore the standard BBN fusion processes
that involve these elements.

To account for bound state effects, an accurate calculation of their abundance
is necessary. To do this we solve numerically  the Boltzmann equations (13) and (14)
from~\cite{kota}, that control these abundances.
If $X$ denotes the light element, and ignoring the fusion contribution as described before, the
system of the  two differential equations for the light-element and bound state
abundances can be cast into the form
\bea
\dot Y_{X}&=& \frac{\left\langle \sigma_c v \right \rangle}{H\, T} \,
                                          (Y_{X} \, n_{\stau} - Y_{BS} \, n^\prime_{\gamma} ) \nonumber \\
\dot Y_{BS}&=& -\dot Y_{X} \, ,
\label{eq:boltzmann}
\eea
where $Y_{BS,X}=n_{BS,X}/s$ and  $n_{\stau}$  is the stau number density.
The thermally-averaged capture cross section 
$\left\langle \sigma_c v \right \rangle $ and 
the photon density $n^\prime_{\gamma}$  for $E> E_{\textrm{bin}} $, are given in Eqs (9)
and (15) in~\cite{kota}, respectively. $H$ is the Hubble expansion and dot denotes derivatives
with respect to the temperature. As initial condition,
we assume that  the bound state abundance is negligible for a temperature of a few times $T_c$. In our
numerical analysis we solve the system (\ref{eq:boltzmann}) for $X=\he4,\li7,\be7$ to
obtain the corresponding $Y_{BS}$ at temperatures below $T_c$.
We assume that the bound state
is destroyed in the reaction.  That is, we do not include additional
bound-state effects on the final-state nuclei such as \li6.

As we see in the following section, bound state effects indeed greatly enhance \li6 production as
found in the analysis of $d(\alpha,\gamma)\li6$ by \cite{maxim}.
Our systematic inclusion of bound state effects finds that \li7 is also significantly altered.
The most important rates are for radiative capture reactions, which
enjoy large boosts due to virtual photon effects.
In particular, bound state \li7 production is dominated by 
the $\iso{H}{3}(\alpha,\gamma)\li7$ and $\he3(\alpha,\gamma)\be7$ rates.
Destruction is dominated by the channel with
the lowest Coulomb barrier, namely $\li7(p,\gamma)2\he4$. Note that \be7 destruction channels
are less important, since mass-7 is largely in \li7 
at $T \ga 60$ keV, and because the high binding energy of [$\stau,\be7$]
makes $[\stau,\be7] + p \rightarrow \iso{B}{8} + \stau$ energetically
forbidden with $Q = - 1.3$ MeV (see Table \ref{tab:enhance}).

Interestingly, the bound state perturbations lead to net \li7 production in
some parts of the parameter space of the  models we study, and net destruction in others.
Net production occurs when the stau is sufficiently
long-lived ($\tau_{\tilde\tau} \ga 10^4$ sec) that \he4
bound states are abundant enough to drive bound state enhanced production
stronger than bound state enhanced \li7 destruction.  On the other hand, 
within a window of slightly
shorter lifetimes, staus persist long enough to form \li7 bound states, 
but then decay before
forming \he4 bound states.  This leads to net \li7 destruction.  Thus
we see that \li7 is quite sensitive to the $\stau$ properties, and
potentially offers a strong probe of the existence and nature of bound
states.

\section{Results and Discussion}

As described earlier, we work in the context of the CMSSM or mSUGRA.
Our primary goal is to examine the effect of bound-state interactions
on the final abundances of the light elements.
To this end, we display a selection of results for specific 
supersymmetric planes both with and without the effect of
bound state interactions.  All results shown fully incorporate the
effects of electromagnetic and hadronic showers.   A more complete selection of 
results as well as constraints which go beyond the MSSM will be
presented in~\cite{cefos2}.  

\begin{figure}
\vskip 0.5in
\vspace*{-0.75in}
\begin{minipage}{8in}
\epsfig{file=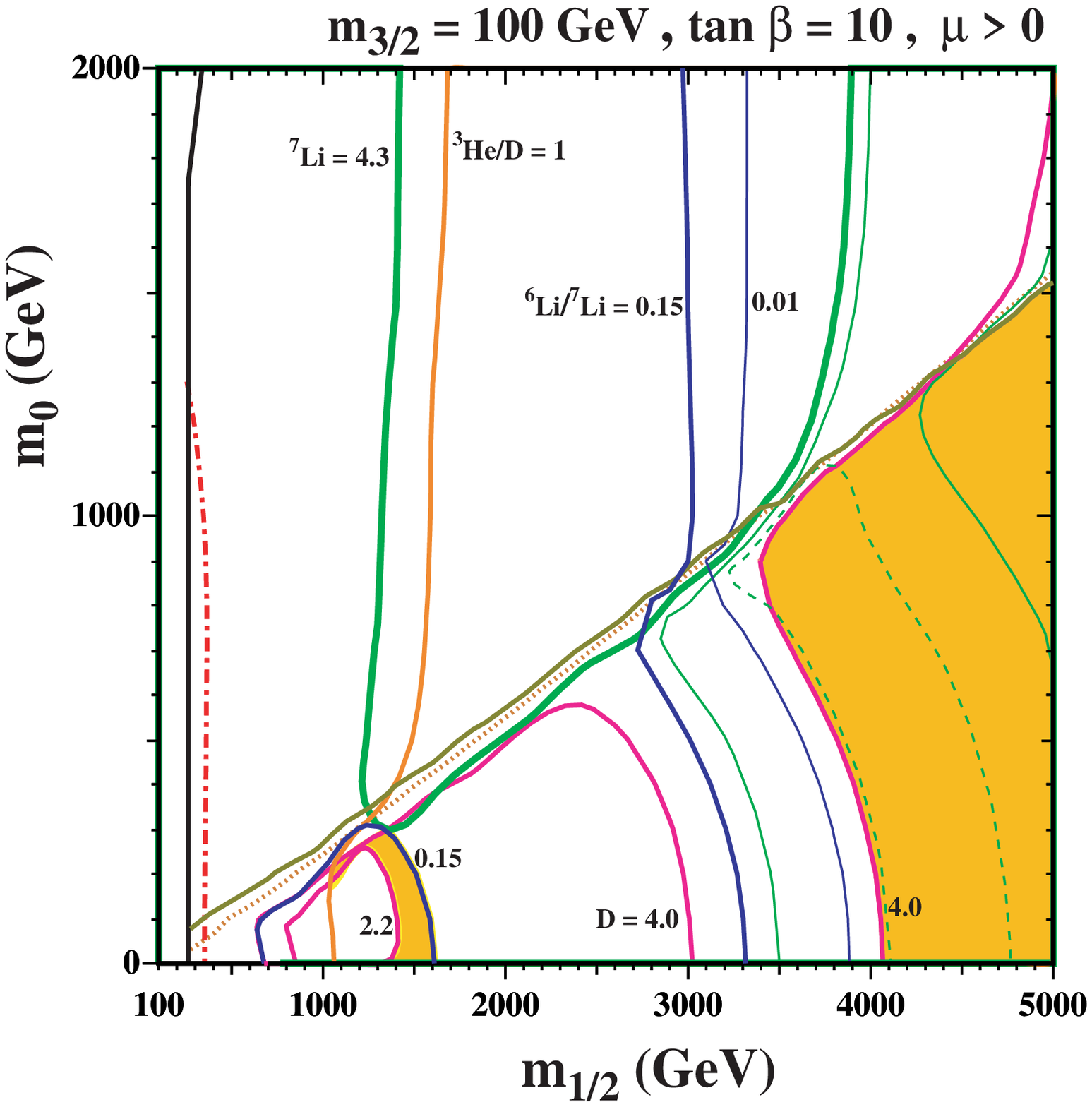,height=3.3in}
\hspace*{-0.17in}
\epsfig{file=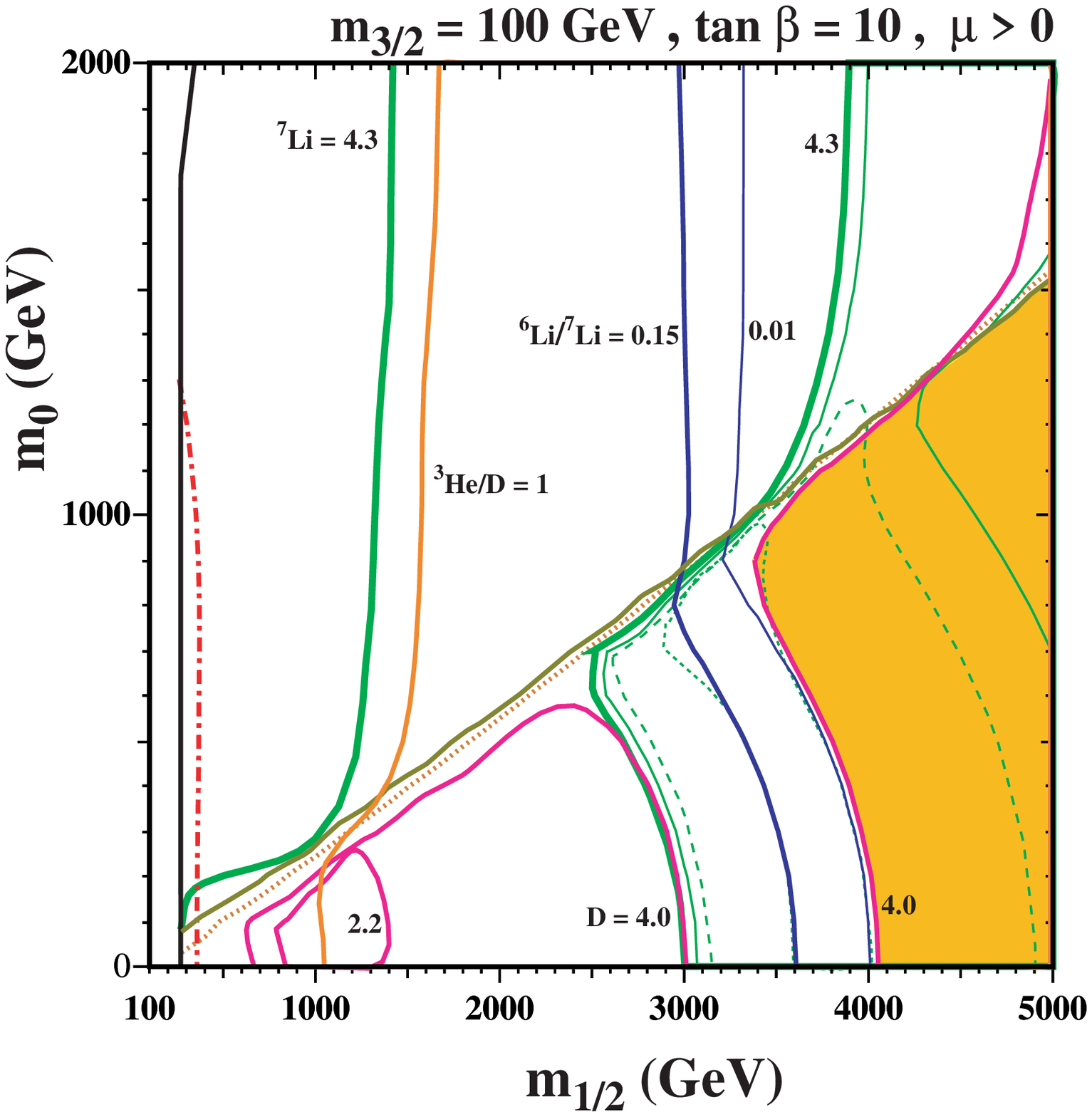,height=3.3in}
\hfill
\end{minipage}
\begin{minipage}{8in}
\epsfig{file=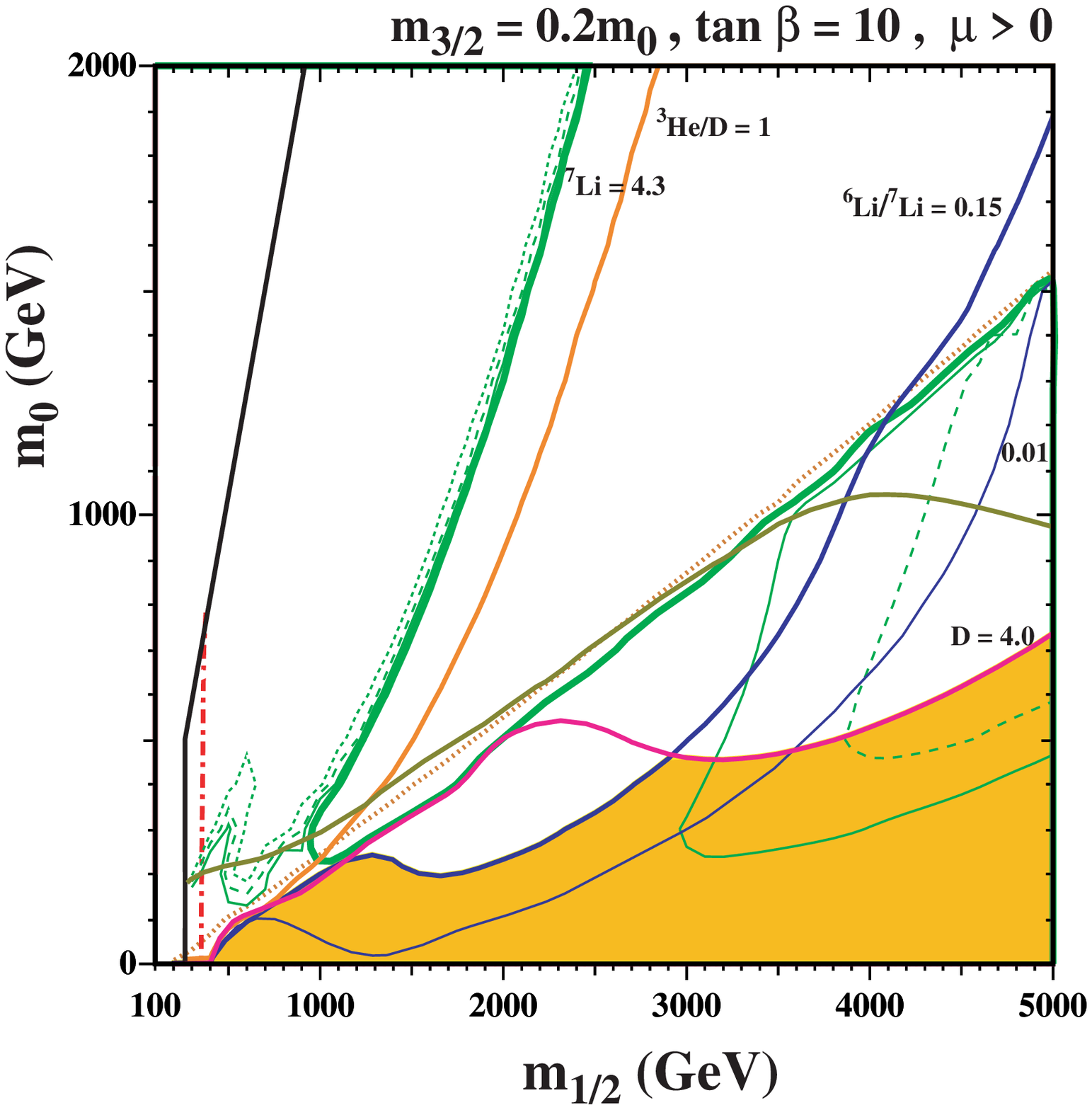,height=3.3in}
\hspace*{-0.2in}
\epsfig{file=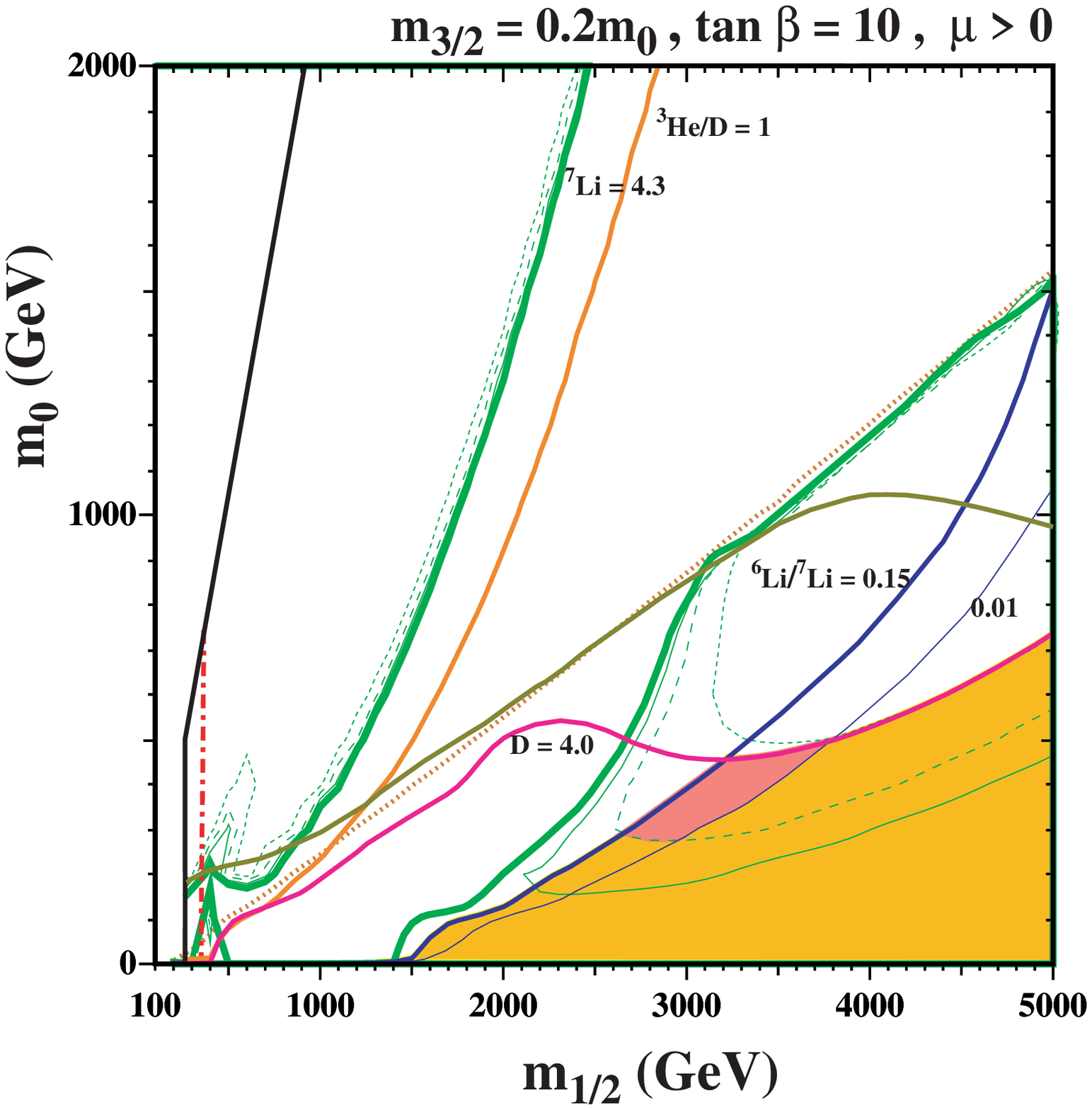,height=3.3in} \hfill
\end{minipage}
\caption{
{\it 
Some $(m_{1/2}, m_0)$ planes for $A_0=0$, $\mu > 0$ and $\tan
\beta = 10$. In the upper (lower) panels we use $m_{3/2} = 100$~GeV
$(m_{3/2} = 0.2\, m_0)$.
In the right panels  the effects of the stau bound states have been included, while 
in those on the left we include only the effect of the NSP decays. 
The regions to the left of the solid black lines are not considered, 
since there the gravitino is not the LSP.
In the orange (light) shaded regions, the differences between the calculated 
and observed light-element abundances are no greater than in standard 
BBN without late particle decays. In the pink (dark) shaded region in 
panel d, the 
abundances lie within the ranges favoured by observation, as described in 
the text. 
The significances of the 
other  lines and contours  are explained in the text.}}
\label{fig:CMSSM10} 
\end{figure}

We begin with results based on CMSSM models with $A_0 = 0$, $\mu > 0$ and
$\tan  \beta = 10$.  We display our results in the $(m_{1/2}, m_0)$ plane,
showing explicit element abundance contours.  
In Fig. \ref{fig:CMSSM10}a, we show the element abundances that result
when the gravitino mass is held fixed at $m_{3/2} = 100$ GeV in the 
{\it absence}
of stau bound state effects.
To the left of the near-vertical solid black line at $m_{1/2} \simeq 250$ GeV,
the gravitino is the not the LSP, and we do not consider this region here.
Immediately to the right of this line is a red dot-dashed line. To the left of this,
the Higgs mass is below the current experimental bound $114\gev$.
The diagonal red dotted line corresponds to the boundary between
a neutralino and stau NSP.  Above the line, the neutralino is the NSP,
and below it, the NSP is the stau.  Very close to this boundary,
there is a diagonal brown solid line.  Above this line, the relic
density of gravitinos from NSP decay is too high, i.e.,
\beq
\frac{m_{3/2}}{m_{NSP}} \Omega_{NSP} h^2 > 0.12.
\eeq
Thus we should restrict our attention to the area below this line.
Note that we display the extensions of contours which originate below the line into the overdense region, but we do not display 
contours that reside solely in the upper plane.

We start with the solid orange line labelled \he3/D = 1.  To the left of this
curve, the \he3/D ratio is greater than 1, which is excluded~\cite{sigl}. 
For small $m_0$, this excludes gaugino masses less than about 1100~GeV,
which is similar to the result found in~\cite{EOV}. To the right of this curve,
the ratio of \he3 to D is acceptable.  The very thick green line labelled \li7 = 4.3
corresponds to the contour where \li7/H = $4.3 \times 10^{-10}$, a value very
close to the standard BBN result for \li7/H. 
It forms a `V' shape, whose right edge runs along the neutralino-stau NSP border
before shooting up at $m_{1/2} \sim 3900$ GeV.
Below the V, the abundance of \li7 is smaller than the standard BBN result. 
However, for relatively small values of $m_{1/2}$, 
the \li7 abundance does not differ very much from this standard BBN result:
it is only when $m_{1/2} \ga 3000$~GeV that \li7 begins to drop significantly.  
This is seen by the additional (unlabeled) thin green contours showing \li7/H = 
$3 \times 10^{-10}$ (solid), and $2 \times 10^{-10}$ (dashed). 
As  can be seen in Fig.~\ref{fig:lifetimes}a, 
the the stau lifetime drops with increasing $m_{1/2}$, and when 
$\tau \sim 1000$ s, at $m_{1/2} \sim 4000$ GeV, the \li7 abundance has been reduced 
to an observation-friendly value close to $2 \times 10^{-10}$ as claimed in~\cite{jed}.

However, for this case with $m_{3/2} = 100$ GeV the \li6 abundance is never sufficiently
high to match the observed \li6 plateau for the same parameter values where \li7 is reduced.
The \li6/\li7 ratio is shown by the solid blue contour labeled \li6/\li7 = 0.15.
Note that there is also a small contour loop at this value at small $m_0$
centred around $m_{1/2} \sim 1000$ GeV.  Inside the loop, the lithium isotope
ratio is acceptable, which is also the case to the right of the nearly vertical contour at large $m_{1/2}$. 
At large $m_{1/2}$, the contour for \li6/\li7 = 0.01 is shown by the thin blue line.
To the right of this contour, including the region where \li7 $\sim 2 \times 10^{-10}$,
the \li6 abundance is too small. 

Finally, we show the contours for D/H = 2.2 and 4.0 $\times 10^{-5}$
by the solid purple contours as labeled.  The D/H = 2.2  $\times 10^{-5}$
contour is a small loop within the \li6/\li7 loop.  Inside this loop
D/H is too small.  Between the two curves labeled 4.0, the D/H ratio is high, 
but not necessarily excessively so. 

In summary, the acceptable regions found in Fig. \ref{fig:CMSSM10}a break
down into 2 areas:  one between the two loops labeled 2.2 and 0.15 and to
the right of the \he3/D = 1 line, where D/H is larger than 2.2 $\times
10^{-5}$ and $\li6/\li7 < 0.15$. However, in this region, the \li7
abundance is very similar to the standard BBN result, which may be
considered too high.  Alternatively, one could consider very large
$m_{1/2}$ where once again D/H $< 4.0 \times 10^{-5}$.  Here, \li7 is in
fact acceptably low, but the \li6 abundance is far below the plateau
value.  As a better illustration of our results, we have shaded these two
regions. The orange (lighter) shaded region is where $\he3/D < 1$,
$\li6/\li7 < 0.15$, $2.2\times 10^{-5} < D/\textrm{H} < 4.0 \times10^{-5}$ and $\li7/\textrm{H}
< 4.3 \times10^{-10} $. 

Turning now to Fig.~\ref{fig:CMSSM10}b, we show the analogous results when
the bound-state effects are included in the calculation.  The abundance
contours are identical to those in Fig.~\ref{fig:CMSSM10}a above the
diagonal dotted line, where the NSP is a neutralino and bound states do
not form.  We also note that the bound state effects on D and \he3 are
quite minimal, so that these element abundances are very similar to those
in Fig.~\ref{fig:CMSSM10}a.  However, comparing panels a and b, one sees
dramatic bound-state effects on the lithium abundances.  The loop of
\li6/\li7 = 0.15 centred about $m_{1/2} = 1000$ GeV has now gone due to
the large abundance of \li6 produced by bound-state catalysis.  Indeed,
everywhere to the left of the solid blue line labeled 0.15 is excluded.  
In the stau NSP region, this means that $m_{1/2} \ga 3000 - 3500$~GeV.  
Moreover, in the stau region to the right of the \li6/\li7 = 0.15 contour,
the \li7 abundance drops below $9 \times 10^{-11}$ (as shown by the thin
green dotted curve) and D/H $> 4\times10^{-5}$ for $m_{1/2} \la 3500 -
4000$~GeV. Only when $m_{1/2} \ga 3500 - 4000$ GeV does the D/H abundance
drop back to acceptable levels with good abundances for \li7, but \li6 is
now too small to account for the plateau.  Thus, for a constant value of
$m_{3/2} = 100$ GeV, the bound-state effects force one to extremely large
values of $m_{1/2}$ primarily due to the enhanced production of \li6, as
shown by the orange shaded region. {\it For this value of the gravitino
mass, there are no regions where both lithium abundances match their
plateau values.}

We do not display the results for $m_{3/2} = 10$ GeV, but the bound-state
effects (and the results) are less dramatic.  Without the bound-state
effects included, the \li6 abundance is generally too small, while the
\li7 abundance is very similar to standard BBN in the stau NSP region. The
gravitino relic density is a factor of 10 smaller in this case and some of
the neutralino NSP region is allowable.  In the neutralino NSP region, D/H
is too high unless $m_{1/2} \ga 2000$ GeV.  At $(m_{1/2},m_0) \simeq
(2100,1000)$, there is a region where D/H and \li7 are acceptably small,
though \li6/\li7 is very small. The bound-state effects again set a lower
limit on $m_{1/2}$ in the stau NSP region in this case. When $m_{1/2} \ga
1300$ GeV, both Li abundances drop and approach their standard BBN values.  
Once again, in no region are both lithium isotopes at their plateau
values.

It is also interesting to consider cases in which the gravitino mass is
proportional to $m_0$. In Fig.~\ref{fig:CMSSM10}c, we fix $m_{3/2} = 0.2
\, m_0$ and neglect the bound-state effects. The choices of contours are
similar to those in panels a and b.  The gravitino relic density
constraint now cuts out some of the stau NSP region at large $m_{1/2}$ and
large $m_0$, but allows a small neutralino NSP region at low $m_{1/2}$. As
before, we are constrained to the right of the curve labeled \he3/D = 1,
though in this case the constraint is not very strong in the stau NSP
region.  The \li7/H = $4.3 \times 10^{-10}$ contour again forms a `V'
shape and one is restricted to lie below the `V'.  In most of the stau NSP
region, \li7 remains relatively high but begins to drop at large $m_{1/2}
\sim 3000$ GeV as $\tau$ approaches $O(1000)$~s (see Fig.~\ref{fig:lifetimes}). 
The region where the
\li6/\li7 ratio lies between 0.01 and 0.15 now forms a band which moves
from lower left to upper right.  Thus, as one can see in the orange
shading, there is a large region where the lithium isotopic ratio can be
made acceptable. However, if we restrict to D/H $< 4.0 \times 10^{-5}$, we
see that this ratio is interesting only when \li7 is at or slightly below
the standard BBN result. However,
we do note that as one approaches the gravitino density limit at
$(m_{1/2}, m_0) \simeq (4100, 1000)$, it is possible to have \li6/\li7
$\simeq 0.06$ and \li7/H $\simeq 2.3 \times 10^{-10}$ at the expense of
D/H $\simeq 6.2 \times 10^{-10}$.

The bound-state effects when $m_{3/2} = 0.2 \, m_0$ are shown in
Fig.~\ref{fig:CMSSM10}d. Once again, we see that the increased production
of both \li6 and \li7 excludes a portion of the stau NSP region where
$m_{1/2} \la 1500$ GeV for small $m_0$.  The lower bound on $m_{1/2}$
increases with $m_0$.  In this case, not only do the bound-state effects
increase the \li7 abundance when $m_{1/2}$ is small (i.e., at relatively
long stau lifetimes), but they also decrease the \li7 abundance when the
lifetime of the stau is about 1500~s. Thus, at $(m_{1/2}, m_0) \simeq
(3200,400)$, we find that \li6/\li7 $\simeq 0.04$, \li7/H $\simeq 1.2
\times 10^{-10}$, and D/H $\simeq 3.8 \times 10^{5}$.  Indeed, when
$m_{1/2}$ is between 3000-4000 GeV, the bound state effects cut the \li7
abundance roughly in half. {\it In the darker (pink) region (which has no 
analogue in
the other panels), the lithium abundances match the observational plateau
values}, with the properties $\li6/\li7 > 0.01$ and $0.9 \times10^{-10} <
\li7/\textrm{H} < 2.0 \times10^{-10}$.

For a larger ratio of $m_{3/2} = m_0$, the gravitino relic density forces
us to relatively low values of $m_0 \la 500$ GeV.  As in the case
described above, the viable region at low $m_{1/2}$ is excluded by the
bound-state effects, and we find increased \li7 production due to bound
states at high $m_{1/2}$.  Qualitatively, this case is similar to that
when $m_{3/2} = 0.2\, m_0$, though most features are compressed to lower
values of $m_0$.

\begin{figure}
\vskip 0.5in
\vspace*{-0.75in}
\begin{minipage}{8in}
\epsfig{file=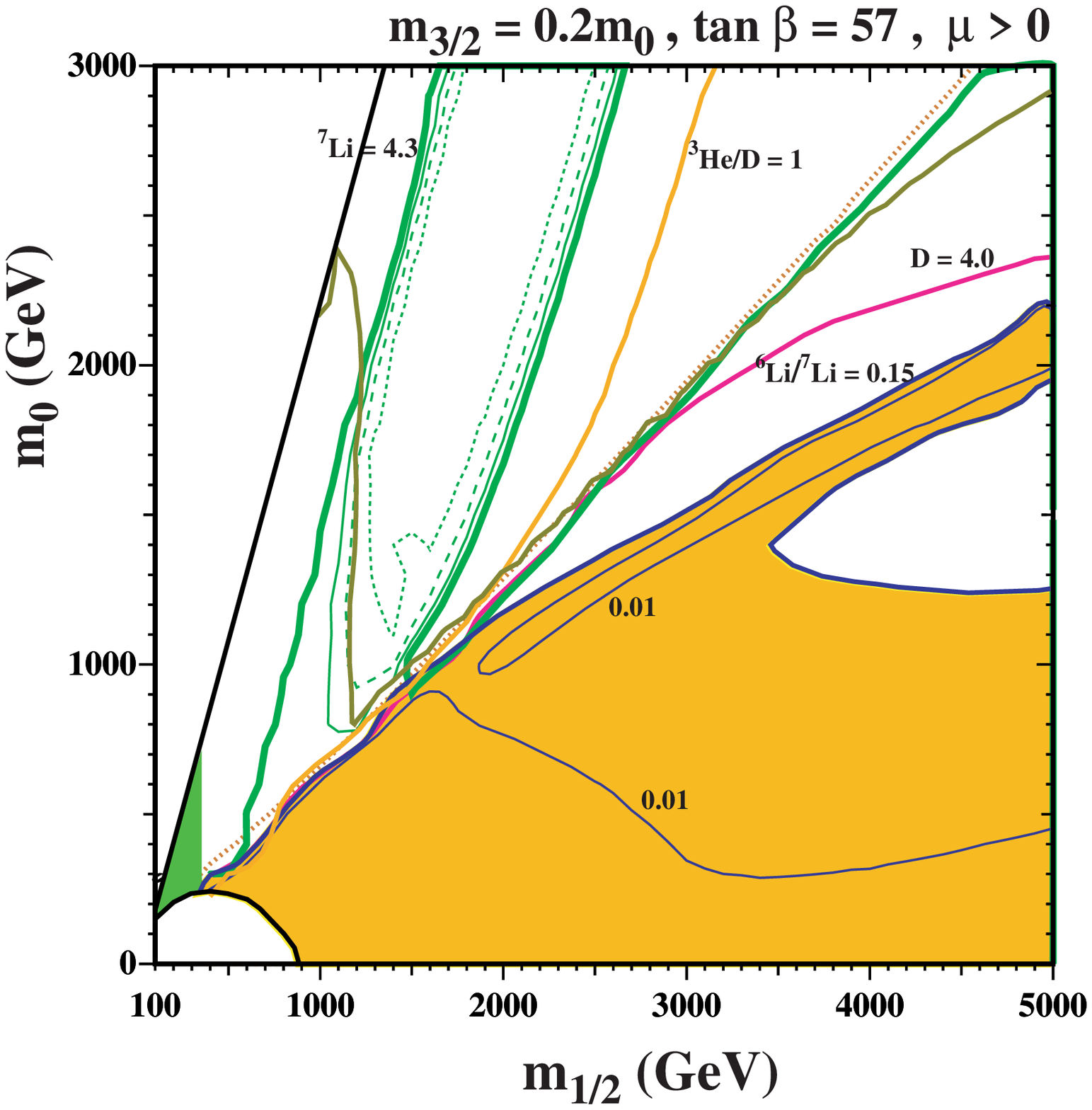,height=3.3in}
\hspace*{-0.17in}
\epsfig{file=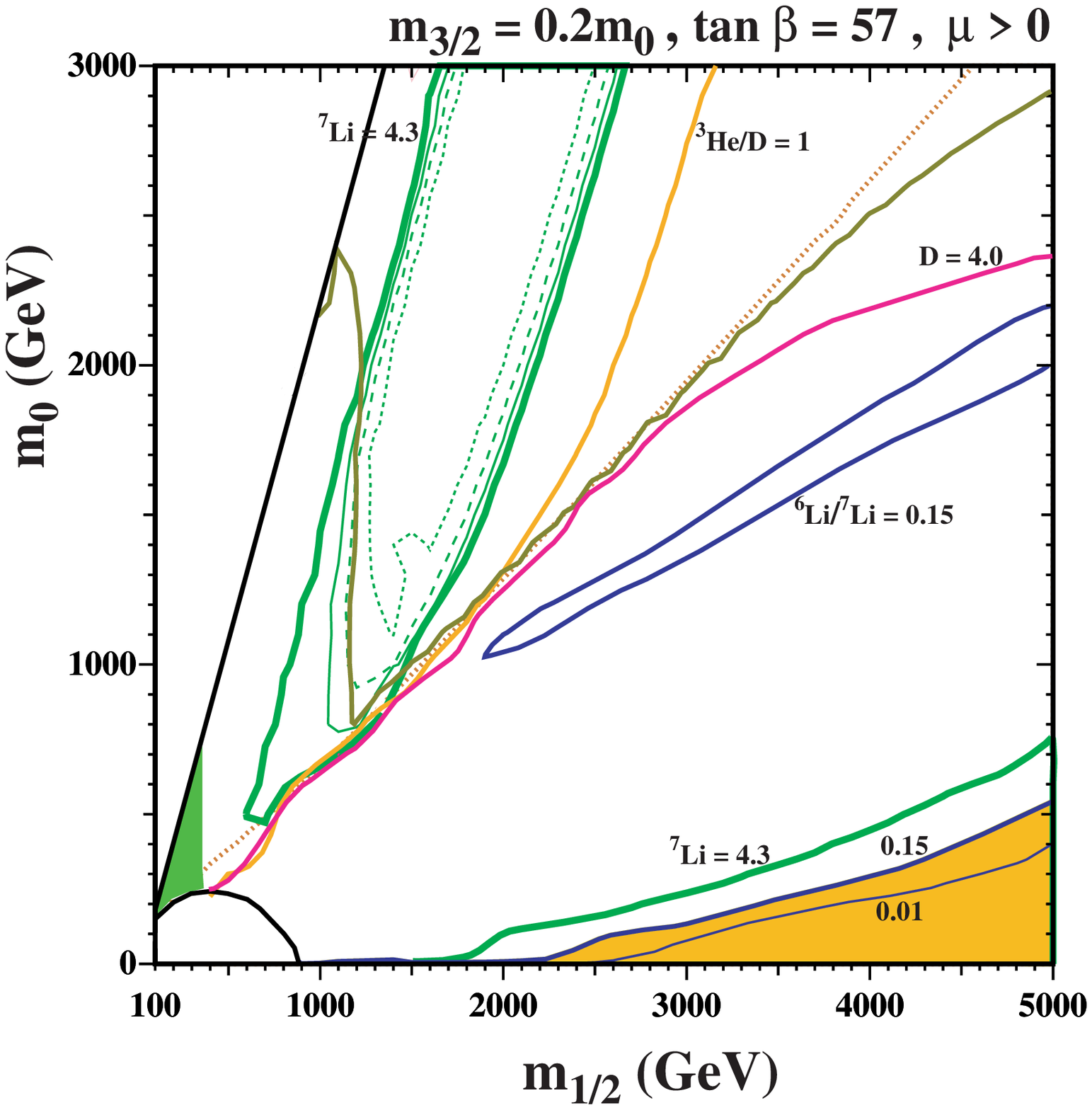,height=3.3in}
\hfill
\end{minipage}
\begin{minipage}{8in}
\epsfig{file=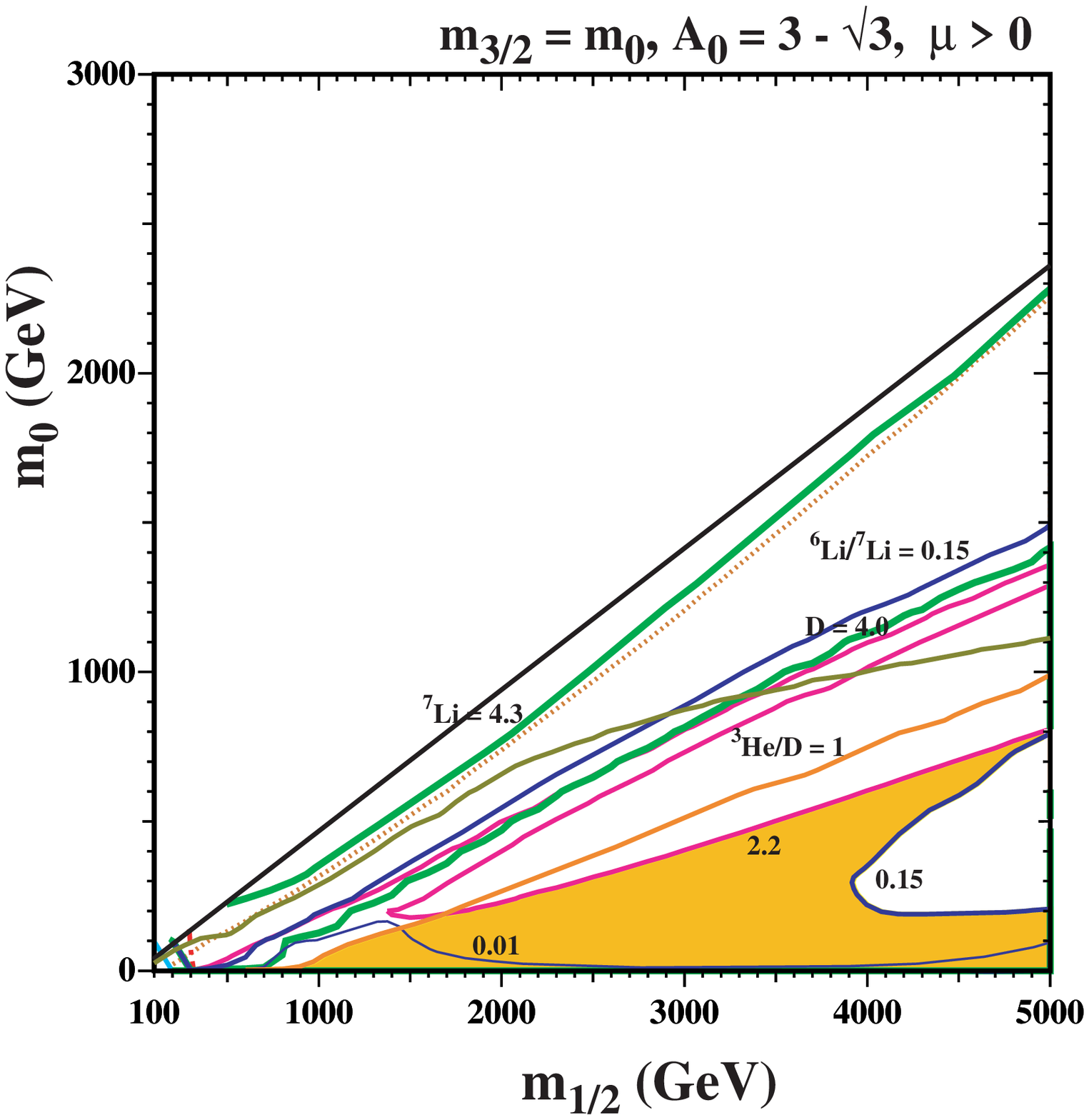,height=3.3in}
\hspace*{-0.2in}
\epsfig{file=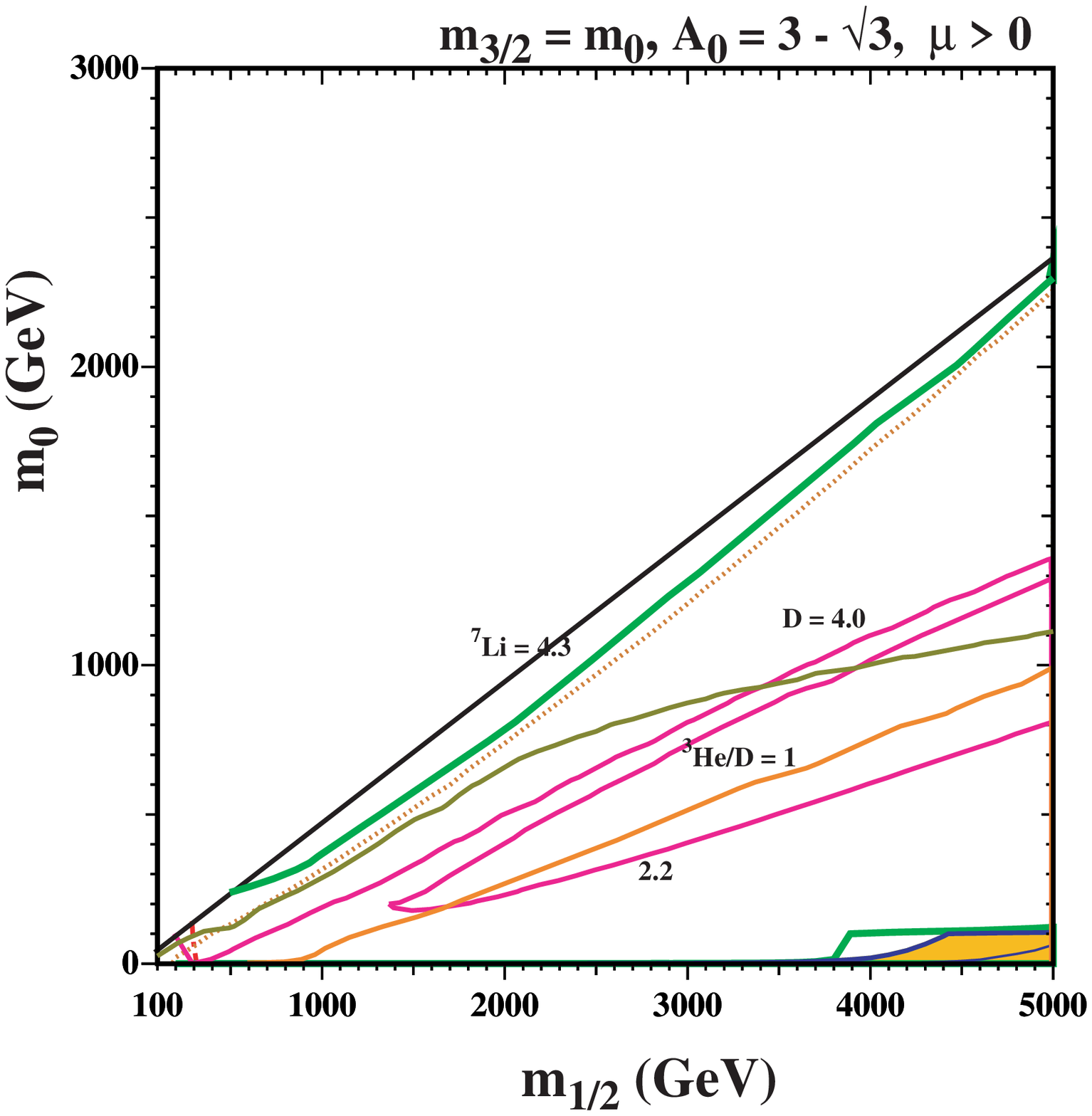,height=3.3in} \hfill
\end{minipage}
\caption{
{\it 
Some more $(m_{1/2}, m_0)$ planes for $\mu > 0$. In the upper panels we
use $m_{3/2} = 0.2\, m_0$ and $\tan \beta = 57$, whilst in the lower
panels we assume mSUGRA with $m_{3/2} = m_0$ and $A_0/m_0=3-\sqrt{3}$ as
in the simplest Polonyi superpotential. In the right panels the effects of
the stau bound states have been included, while in those on the left we
include only the effects of the NSP decays.  As in Fig.~\ref{fig:CMSSM10},
the region above the solid black line is excluded, since there the
gravitino is not the LSP.
In the orange shaded regions, the differences between the calculated 
and observed light-element abundances are no greater than in standard 
BBN without late particle decays. The meanings of the other lines and contours
 are explained in the text.}}
\label{fig:CMSSM57} 
\end{figure}

In Fig.~\ref{fig:CMSSM57}, we show some examples of results from CMSSM
models with $\tan \beta = 57$, and mSUGRA models.  The dominant effect of
increasing $\tan \beta$ is on the neutralino and stau relic densities. At
low $\tan \beta$, the relic density of the neutralino is generally high
except along a narrow strip where neutralino-stau co-annihilations are
important and yield a density with in the WMAP range.  At large $\tan
\beta$, new annihilation channels are available. Most predominant is the
the s-channel annihilation of neutralinos through the heavy Higgs scalar
and pseudoscalar, causing large variations in the relic density across the
plane, particularly at large $m_{1/2}$ and $m_0$.  These variations have
an impact in GDM scenarios, as the abundance of decaying particles varies.

In Fig.~\ref{fig:CMSSM57}a, we show the $(m_{1/2}, m_0)$ plane for $\tan
\beta = 57$ (which is near the maximal value for which the electroweak
symmetry breaking conditions can be satisfied), and $m_{3/2} = 0.2\, m_0$.  
The dark green shaded region at very low $m_{1/2}$ is excluded by
$b \to s \gamma$ decays.
Notice that the constraint on the gravitino relic density (shown by the
solid brown line) no longer tracks the neutralino-stau NSP border.  At
$m_{1/2} \sim 1200$ GeV, it shoots upwards towards large $m_0$.  This is
due to the s-channel annihilation pole (where $2m_\chi = m_A$) which
decreases the relic density.  Consider now the behavior of the \li7
abundance as $m_{1/2}$ is increased at a fixed value of $m_0 = 2000$ GeV.  
At small $ m_{1/2} \la 1000$ GeV, the neutralino is the LSP and results
are not shown.  When $\tan \beta = 10$, the relic neutralino density is very
large, and when $m_{1/2} \la 2400$ GeV the lifetime is greater than $5
\times 10^4$ s, and \li7 destruction process are very efficient.  At
larger $m_{1/2}$, the lifetime decreases and hadronic production effects
begin to dominate, and the \li7 abundance becomes very large.  When $\tan
\beta = 57$, the relic density of neutralinos is 2-3 orders of magnitude
smaller when $m_{1/2} \la 1500$ GeV, for the same value of $m_0$.  As a
result, \li7 destruction is suppressed.  As $m_{1/2}$ is increased, and we
move away from the pole, the neutralino density increases, dropping the
\li7 abundance for long lifetimes. As for lower $\tan \beta$, as $m_{1/2}$
is further increased and the $\tau_\chi$ decreases, the \li7 abundance
becomes large again, until one hits the neutralino-stau NSP border.  The
`V'-shaped \li7 contour at large $m_{1/2}$ is visible here as well, though
it appears squeezed as the NSP border is moved up at large $\tan \beta$.

Just below the NSP border, we see another distinctive feature in
Fig.~\ref{fig:CMSSM57}a. The \li6/\li7 ratio, which is generally too large
when the neutralino is the NSP, drops dramatically inside a narrow
diagonal strip.  This occurs because the annihilations of staus are here
dominated by a similar s-channel pole.  Inside this strip, the density of
staus is very small, and element abundances approach their standard BBN
values.  At lower $m_0$, over much of the plane with a stau NSP, the \li7
and D abundances are close to their standard BBN values, while \li6 is
enhanced. In this case, there is a substantial orange shaded region where 
the light-element abundances are no less acceptable than in standard BBN.

Our results for $\tan \beta = 57$ and $m_{3/2} = 0.2 \, m_0$ when the
bound-state effects are included are shown in Fig.~\ref{fig:CMSSM57}b.  
For $\tan \beta = 57$, the stau lifetimes are somewhat longer than the
corresponding lifetimes when $\tan \beta = 10$.  This means that the
bound-state effects are apparent over a larger portion of the plane with a
stau NSP. Both lithium isotope abundances are significantly higher.
Without the bound states, the \li7 abundance varies little from its
standard BBN value, but with their inclusion the \li7 abundance is
somewhat higher, generally about $5 \times 10^{-10}$. The effect on \li6
is larger.  Without the bound-state effects, the \li6/\li7 ratio remains
small unless either $m_{1/2}$ and/or $m_0$ are relatively large. Even
then, the ratio only increases to a few percent unless $m_{1/2} \sim 3500
- 5000$ GeV and $m_0 \sim 1200 - 1900$ GeV, where it exceeds 0.15.  With
the inclusion of the bound states, in much of the stau NSP region the \li6
abundance is too high, exceptions being the area where s-channel
annihilation occurs or in the lower right corner of the displayed plane.
{\it There is no region where the light-element abundances lie in the 
favoured plateau ranges.}

Finally, we come to an example of a mSUGRA model.  Here, because of a
relation between the bilinear and trilinear supersymmetry breaking terms:
$B_0 = A_0 - m_0$, $\tan \beta$ is no longer a free parameter of the
theory, but instead must be calculated at each point of the parameter
space.  Here, we choose an example based on the Polonyi model for which
$A_0/m_0 = 3- \sqrt{3}$.  In addition, we have the condition that $m_{3/2}
= m_0$.  In Fig.~\ref{fig:CMSSM57}c, we show the mSUGRA model without the
bound states. In the upper part of the plane, we do not have GDM.  We see
that \he3/D eliminates all but a triangular area which extends up to $m_0
= 1000$ GeV, when $m_{1/2} = 5000$ GeV.  Below the \he3/D = 1 contour, D
and \li7 are close to their standard BBN values, and there is a
substantial orange shaded region. We note that \li6 is interestingly high,
between 0.01 and 0.15 in much of this region.

As seen in Fig.~\ref{fig:CMSSM57}d, when bound-state effects are included
in this mSUGRA model, both lithium isotope abundances are too large except
in the extreme lower right corner, where there is a small region shaded 
orange. {\it However, there is no region where the lithium abundances fall 
within the favoured plateau ranges.}

\section{Conclusions}

We have calculated in this paper the cosmological light-element abundances
in the presence of the electromagnetic and hadronic showers due to late
decays of the NSP in the context of the CMSSM and mSUGRA models,
incorporating the effects of the bound states that would form between a
metastable stau NSP and the light nuclei.  Late decays of the neutralino
NSP constrain significantly the neutralino region, since in general they
yield large light-element abundances.  The bound-state effects are
significant in the stau NSP region, where excessive \li6 and \li7
abundances exclude regions where the stau lifetime is longer than
$10^3-10^4$~s. For lifetimes shorter than $1000$~s, there is a possibility
that the stau decays can reduce the \li7 abundance from the standard BBN
value, while at the same time enhancing the \li6 abundance. A more 
complete account of our calculations will be given in~\cite{cefos2}, where 
more examples of CMSSM and mSUGRA parameter planes will be presented, and 
the possibility of matching the favoured lithium abundances will 
be discussed in more detail.
 
\section*{Acknowledgments}
\noindent 
We would like to thank M. Pospelov and M. Voloshin for helpful discussions.
The work of K.A.O. and V.C.S. was supported in part
by DOE grant DE--FG02--94ER--40823.


\end{document}